\documentclass[aps,prb,reprint,titlepage]{revtex4-2}
\usepackage{amsmath}  
\usepackage{amsfonts} 
\usepackage{graphicx} 
\usepackage[T1]{fontenc}
\usepackage{braket}
\begin{document}

\title{Quantitative Analysis of Exciton Composition and Dynamics in Y6 Films for Single-Component Solar Cells}

\author{Saba Mahmoodpour}
\affiliation{Department of Chemistry, University of North Carolina at Chapel Hill, Chapel Hill, NC 27599, USA}
\author{Jiyeon Oh}
\affiliation{Department of Chemistry, University of North Carolina at Chapel Hill, Chapel Hill, NC 27599, USA}
\author{Yanlin Liu}
\affiliation{Department of Chemistry, University of North Carolina at Chapel Hill, Chapel Hill, NC 27599, USA}
\author{Zijian Gan}
\affiliation{Department of Chemistry, University of North Carolina at Chapel Hill, Chapel Hill, NC 27599, USA}
\author{Wei You}
\affiliation{Department of Chemistry, University of North Carolina at Chapel Hill, Chapel Hill, NC 27599, USA}
\author{Andrew M. Moran}
\email{ammoran@unc.edu}
\affiliation{Department of Chemistry, University of North Carolina at Chapel Hill, Chapel Hill, NC 27599, USA}


\begin{abstract}
Non-fullerene acceptors such as Y6 have enabled high-efficiency organic photovoltaic devices and motivated the development of single-component architectures; however, the microscopic mechanisms governing exciton transport and charge dissociation remain under active investigation. In particular, the interplay between Frenkel and charge--transfer excitations and their coupling to environmental fluctuations complicates the description of light absorption and subsequent exciton dynamics. Here, ultrafast transient absorption spectroscopy is used to probe exciton quenching dynamics in Y6 films interfaced with hole-transport layers. To interpret these measurements, we develop an analytical model based on hybrid Frenkel--charge--transfer states that enables direct extraction of intermolecular electronic couplings, charge--transfer character, and system--bath interaction strengths from experimental data. The analysis reveals a substantial charge--transfer admixture of 20--40\% in the exciton states and identifies a transport regime characterized by delocalization-mediated exciton motion rather than purely diffusive hopping. Consistent with this interpretation, the corresponding quenching dynamics occur on a $\sim$1~ps timescale within $\sim$4~nm of the interface, suggesting a short-range injection mechanism facilitated by exciton delocalization. In addition to providing physical parameters for Y6, these results establish a quantitative framework that connects spectroscopic observables to microscopic transport mechanisms and can be generalized to other non-fullerene acceptors.
\end{abstract}

\maketitle

\newpage 

\section{Introduction}

\begin{figure}[ht]
\centering
\includegraphics[width=1\linewidth]{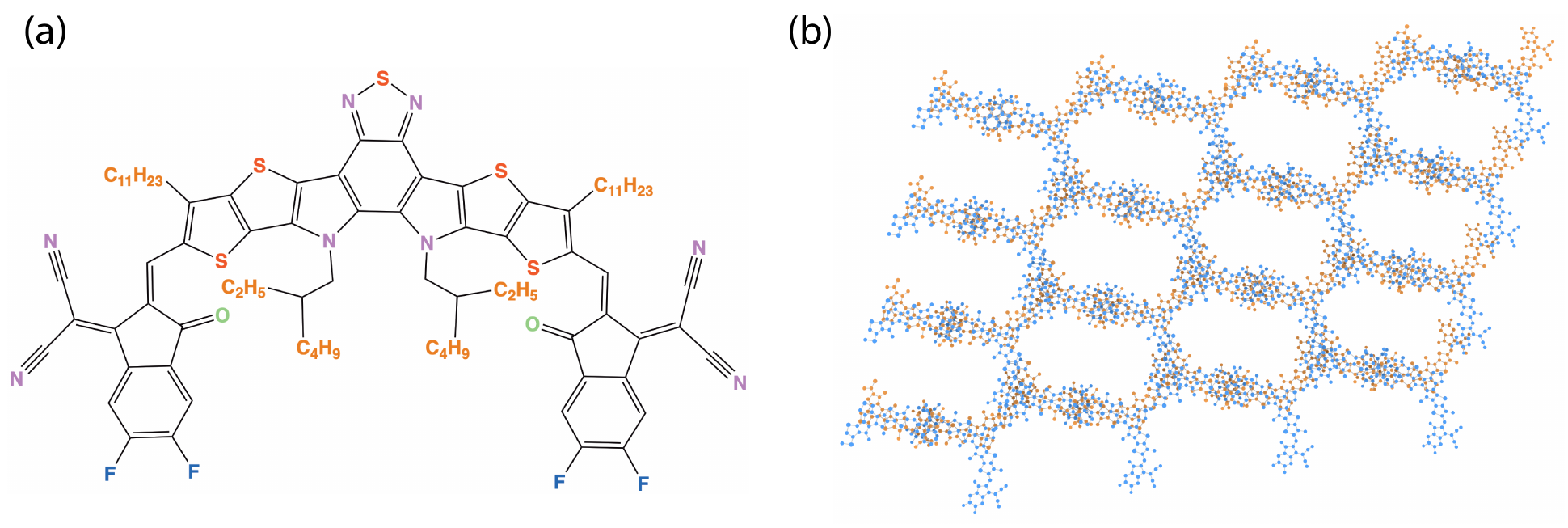}
\caption{Structural representation of Y6 and its packing geometry. (a) Molecular structure of Y6. (b) Top-down view of the cofacial packing geometry adopted by Y6 films under the present fabrication conditions.}
\label{fig:y6structure}
\end{figure}

Organic photovoltaic devices have undergone rapid advances in power conversion efficiency, largely driven by the development of non-fullerene acceptors, with Y6 and its derivatives playing a central role \cite{houOrganicSolarCells2018,yuanSingleJunctionOrganicSolar2019,humeNewAvenuesOrganic2024}. Compared to fullerene-based systems, Y6 exhibits a combination of favorable photophysical properties, including strong near-infrared absorption, low energetic disorder, high charge-carrier mobility, and pronounced intermolecular electronic coupling arising from its packing motifs \cite{xiaoSinglecrystalFieldeffectTransistors2020,perdigón-toroBarrierlessFreeCharge2020,humeNewAvenuesOrganic2024,stojanovićDisorderInducedTransitionTransient2024,westbrookSolidStatePackingControls2025,guoBalancedIntraInterchain2026}. These features have motivated interest in single-component Y6 devices, which provide a simplified platform for promoting fundamental photophysical processes without the need to optimize donor–acceptor energetics and morphology in bulk heterojunctions. Despite this progress, the microscopic nature of the primary photoexcitations and their role in photocurrent generation remain under active debate. A range of studies have proposed that light absorption leads to the formation of FE or CT-like states, with differing interpretations regarding whether the primary photoexcitations are free carriers, bound polaron pairs, or hybridized FE--CT excitations  \cite{zhangDelocalizationExcitonElectron2020,zhuSmallExcitonBinding2021,priceFreeChargePhotogeneration2022,saglamkayaWhatSpecialY62023,mahadevanAssessingIntraIntermolecular2024,cerdáTuningExcitonDiffusion2025,hartMolecularFactorsControlling2026}. However, recent experimental and theoretical studies suggest that significant activation barriers and exciton binding energies persist, implying that excitations remain largely bound in the bulk, where charge separation is strongly influenced by interfaces and molecular packing (Fig.~\ref{fig:y6structure}) \cite{kafleExcitonHeatingCooling2021,xieWhyDoesY62024,gianniniRoleChargeTransfer2024,kumarMorphologicalControlY62025}. These considerations underscore the need for a quantitative framework that connects the nature of the primary excitation to charge transport and extraction in single-component Y6 devices.

In this work, we use transient absorption spectroscopy to investigate exciton dynamics at Y6/hole-transport-layer interfaces and resolve the microscopic transport mechanisms preceding charge dissociation. While conventional fluorescence quenching measurements provide estimates of exciton diffusion lengths \cite{firdausLongrangeExcitonDiffusion2020,logerfomSpatiotemporalMappingUncouples2023}, they do not directly access the relevant timescales. By contrast, transient absorption measurements enable determination of exciton migration dynamics and quenching yields, allowing us to distinguish between diffusive and delocalization-mediated transport within a Redfield framework \cite{BreuerPetruccione2002,yangInfluencePhononsExciton2002,MayKuhn2011,balzerMechanismDelocalizationEnhancedExciton2023}. To interpret these results, we develop a reduced analytical model based on electronic states with mixed FE--CT character that captures the key features of the electronic structure while remaining sufficiently tractable for direct fitting of experimental spectra \cite{hestandExpandedTheoryJMolecular2018,gianniniRoleChargeTransfer2024,cerdáTuningExcitonDiffusion2025}. Fitting the experimental data yields the CT character of the excitations, the redistribution of oscillator strength among exciton states, and the reorganization energies of the FE and CT configurations. Together, these experimental results and modeling approaches establish a direct connection between spectroscopic observables and exciton transport mechanisms, providing a general framework applicable to other non-fullerene acceptor systems.

The optical responses of molecular aggregates and crystals are typically described within a FE framework in which the $\pi$--$\pi^*$ excitations of the individual molecules delocalize over multiple sites through transition dipole couplings \cite{aleksandrsergeevichdavydovTheoryMolecularExcitons1971,agranovichGalanin1982,vanamerongenPhotosyntheticExcitons2000,fassioliPhotosyntheticLightHarvesting2014}. Such delocalization gives rise to collective optical phenomena, including superradiance and the formation of J- and H-aggregates, wherein the absorptivity concentrates at the bottom or top of the FE band \cite{spanoSuperradianceMolecularAggregates1989,knoesterExcitonSuperradianceMolecular1992}. In planar conjugated systems such as Y6 films, cofacial $\pi$–$\pi$ stacking leads to electronic coupling via orbital overlap, resulting in charge-transfer configurations with the electron and hole residing on neighboring molecules. These CT states hybridize with the optically active FE configurations and form mixed excitations whose properties depend on the electronic coupling strength and energetic detuning from the $\pi$--$\pi^*$ resonances \cite{bardeenStructureDynamicsMolecular2014,hestandExpandedTheoryJMolecular2018,ghoshExcitonsPolaronsOrganic2020,gianniniRoleChargeTransfer2024,cerdáTuningExcitonDiffusion2025}. Charge--transfer hybridization redistributes oscillator strength, modifies absorption lineshapes, and reduces exciton binding energies. Additionally, stronger coupling to thermal fluctuations in the surrounding environment promotes transient localization through dynamic disorder, reducing the spatial extent of the hybridized excitons and suppressing delocalization-mediated transport \cite{hakenStrobl1973,ripsStochasticModelsExciton1993,troisiChargeTransportRegimeCrystalline2006,ciuchiTransientLocalizationCrystalline2011}. Consequently, the interplay between FE and CT character governs both the optical response and exciton transport in Y6 and related systems.

Building on these concepts, the reduced FE--CT model developed in the present work treats both the optical response and exciton transport using a minimal set of shared parameters that can be constrained by fitting spectroscopic signals. Steady-state absorbance lineshapes reflect the degree of intensity borrowing via CT mixing as well as the reorganization energies associated with system--environment coupling. Ultrafast quenching measurements provide a complementary perspective, wherein excited-state hybridization influences both the exciton size and the efficiency of energy dissipation accompanying transport to the hole-transfer interface \cite{kafleRelationshipCoherentSize2018}. These competing effects give rise to a turnover in the Redfield exciton transfer rate, indicating that an intermediate degree of CT character is optimal for exciton transport and charge extraction. Fits to the steady-state and transient spectra place the exciton states in this intermediate regime, yielding a CT admixture of $\lesssim50\%$ and delocalization-mediated exciton migration over $\sim4$~nm on a $\sim1$~ps timescale near the hole-transfer interface.

\section{Experimental Methods}
\subsection{Film Fabrication}

The non-fullerene acceptor Y6 was purchased from 1-Material Inc. and used as received without further purification. Anhydrous chloroform, copper(I) thiocyanate (CuSCN), and diethyl sulfide (DES) were purchased from Sigma-Aldrich. Glass substrates were sequentially cleaned by rinsing with detergent, acetone, and isopropanol for 15~min each, followed by ultraviolet--ozone plasma treatment for 15~min.

A CuSCN solution (35~mg~mL$^{-1}$) was prepared by dissolving CuSCN powder in diethyl sulfide (DES) at 60~$^\circ$C for 1~h and subsequently filtered. CuSCN films were deposited onto cleaned glass substrates by spin-coating at 2000~rpm for 30~s, followed by annealing at 100~$^\circ$C for 5~min.

Y6 was dissolved in anhydrous chloroform at various concentrations inside a nitrogen-filled glovebox. The solutions were heated at 65~$^\circ$C with stirring for 120~min prior to deposition. Film thickness was controlled by adjusting both the solution concentration and spin-coating speed. Specifically, $\sim$20~nm films were obtained from a 3~mg~mL$^{-1}$ solution (9000~rpm), $\sim$40~nm films from a 10~mg~mL$^{-1}$ solution (3500~rpm), $\sim$65~nm films from a 14~mg~mL$^{-1}$ solution (3500~rpm), and $\sim$100~nm films from an 18~mg~mL$^{-1}$ solution (2500~rpm). Films were deposited onto cleaned glass substrates by spin-coating under these conditions, followed by thermal annealing at 100~$^\circ$C for 10~min in a nitrogen-filled glovebox.

The Y6 films prepared under these conditions exhibit red-shifted and broadened absorption spectra relative to solution, indicative of intermolecular electronic coupling and cofacial $\pi$--$\pi$ interactions in the solid state \cite{yuanSingleJunctionOrganicSolar2019,zouInsightExcitationStates2020}. These spectral features are consistent with cofacial nearest-neighbor packing and associated H-type (positive) transition dipole couplings between neighboring molecules in solution-cast films \cite{kumarMorphologicalControlY62025}. Prior studies have shown that chloroform processing can favor a face-on orientational bias in Y6 thin films \cite{fuMolecularOrientationdependentEnergetic2023}; however, the formation of cofacially $\pi$-stacked aggregates is the primary structural feature relevant to the excitonic behavior considered here. 

Film thicknesses were independently verified by scanning electron microscopy (SEM) and correlated with optical absorbance to determine the absorption coefficient ($\alpha = 1.14 \times 10^{5}$~cm$^{-1}$), consistent with reported values for Y6 \cite{yuanSingleJunctionOrganicSolar2019,zhangBoostedEfficiency1812022}. This absorption coefficient was used to determine the thicknesses of additional Y6 films.

\subsection{Spectroscopic Measurements}
Transient absorption measurements were performed using a Coherent Libra Ti:sapphire amplifier system that delivers 50-fs pulses at 800~nm with a pulse energy of 3.5~mJ and a repetition rate of 1~kHz. Approximately 1.8~mJ of the fundamental output was focused into a 4-m-long gas cell filled with argon gas at 2.0~atm to produce a continuum via filamentation. Pump pulses centered at either 740 or 795~nm with full widths at half maximum (FWHM) of $\sim$10~nm were selected from the continuum using a home-built 4$f$ spectral filtering apparatus. Broadband probe pulses were generated by focusing a portion of the beam into a sapphire window and delivered to the sample using reflective optics.

The pump and probe beams were focused to spot sizes of approximately 350~$\mu\mathrm{m}$ at the sample position, with pump fluences in the range of 5--10~$\mu\mathrm{J/cm^2}$. Samples were mounted on a continuously rotating stage to minimize photodegradation, and no changes in the dynamics or spectral lineshapes were observed over the 30-minute acquisition time. The CuSCN layer was confirmed not to exhibit a measurable transient absorption response at the pump and probe wavelengths when deposited on substrates without Y6 films. Signals were detected using a CMOS array detector synchronized to the 1-kHz laser repetition rate. Data were averaged over 15 scans of the delay stage, corresponding to a total of $1.2\times10^4$ pump--probe measurements at each delay time.

Two pump--probe configurations were employed. First, to verify consistency with previously reported transient absorption spectra, measurements on neat Y6 films were performed over the full 500--1000~nm spectral range using 740~nm pump pulses, with the dispersed probe intensity balanced by a combination of dielectric mirrors and spectral filters. Second, comparative measurements on samples with and without hole-transport layers (HTLs) were performed using a 795~nm pump wavelength and a restricted probe range of 740--840~nm to study quenching dynamics under identical conditions. For all measurements, the probe intensity was flattened by adjusting the angle of incidence on a dielectric mirror placed before the sample to suppress nonlinear contributions from the fundamental beam. The probe pulses were not compressed, as the $\sim 200$~fs instrument response is sufficient to resolve the quenching dynamics.

\begin{figure}[ht]
\centering
\includegraphics[width=1.0\linewidth]{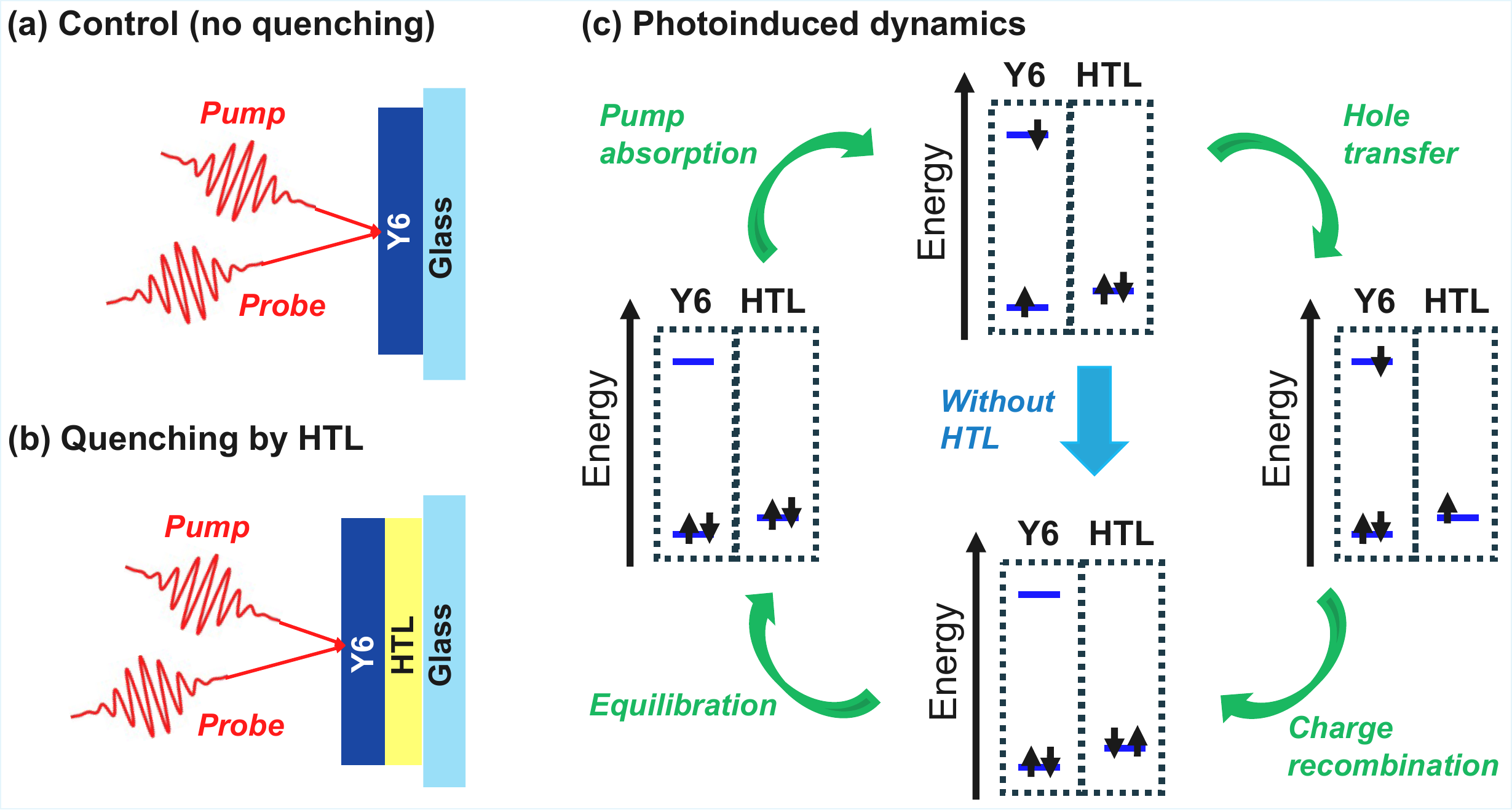}
\caption{Exciton quenching dynamics are investigated by comparing transient absorption measurements on Y6 films (a) without and (b) with the copper(I) thiocyanate hole-transport layer (HTL). (c) Photoexcitation generates excitons that undergo transport and subsequent interfacial hole transfer to the HTL on a picosecond timescale, followed by charge recombination and recovery of the ground state on nanosecond timescales.}
\label{fig:tascheme}
\end{figure}

The experimental approach used to determine the quenching dynamics is illustrated schematically in Fig.~\ref{fig:tascheme}. Photoexcitation of the Y6 film generates excitons that migrate toward the adjacent HTL on a picosecond timescale. In the absence of the HTL, excitons decay through intrinsic relaxation pathways, whereas in the presence of the HTL, interfacial hole transfer provides an additional quenching channel. Measurements on samples with and without the HTL were performed sequentially under identical optical conditions to enable direct comparison of the transient absorption dynamics. To resolve both the ultrafast quenching dynamics and long-time recovery, the delay was scanned over a 5~ns range using logarithmically spaced time steps. The interval between adjacent delay points increased from approximately 60~fs near time zero to 800~ps at $\tau=5$~ns.

Complementary transient absorption anisotropy measurements were also performed on Y6 solutions and films using the same pump--probe apparatus to investigate the exciton electronic structure. Because these measurements provide independent evidence for exciton delocalization but are ancillary to the primary transport analysis, the experimental procedure and data analysis are presented in the Supplemental Material \cite{SM_Y6}.

\section{Frenkel--Charge--Transfer Hybridization in Spectroscopy and Dynamics}
\label{sec:theoryintro}

The electronic structure of molecular semiconductors is conveniently described in a local diabatic basis comprising FE configurations, corresponding to $\pi$--$\pi^*$ transitions localized on individual molecules, and CT configurations with electron--hole separation across neighboring molecules. In Y6 and related systems, strong hybridization between these configurations gives rise to excitonic states with mixed FE--CT character \cite{bardeenStructureDynamicsMolecular2014,hestandExpandedTheoryJMolecular2018,gianniniRoleChargeTransfer2024,cerdáTuningExcitonDiffusion2025,hartMolecularFactorsControlling2026}. Because CT configurations typically exhibit stronger coupling to environmental fluctuations, the degree of CT admixture plays a central role in determining both optical response and relaxation dynamics. These considerations motivate the development of a reduced analytical model whose objective is to retain the key microscopic interactions governing the spectroscopic observables while reducing the complexity to permit quantitative fitting of experimental data. In the following sections, this framework is used to describe the linear absorption spectrum and Redfield population relaxation, providing an integrated description of spectroscopic line shapes and exciton transport.

The nature of the primary photoexcitations in Y6 films remains under active discussion, with some studies proposing efficient bulk generation of free charge carriers \cite{zhuSmallExcitonBinding2021,priceFreeChargePhotogeneration2022,saglamkayaWhatSpecialY62023}, whereas others support bound excitations with mixed FE--CT character \cite{zhangDelocalizationExcitonElectron2020,gianniniRoleChargeTransfer2024,mahadevanAssessingIntraIntermolecular2024,cerdáTuningExcitonDiffusion2025}. In the present work, we adopt the latter description and model transport in terms of partially delocalized FE--CT excitations. This choice is motivated by this body of recent experimental and theoretical work, together with the spectroscopic results presented in this work.

\subsection{Electronic Hamiltonian in the Mixed Basis}
\label{sec:fect_hamiltonian}

We introduce a microscopic Hamiltonian that provides the physical basis for the model used to fit the experimental data. In our notation, an FE basis state, denoted $|n\rangle$, represents a neutral excitation localized on molecule $n$. In contrast, a CT state, denoted $|n m\rangle$, represents a configuration in which a hole resides on molecule $n$ and an electron resides on molecule $m$. The electronic Hilbert space comprises all localized FE states together with the intermolecular CT configurations. Although the CT basis formally includes all intermolecular electron--hole configurations, nearest-neighbor interactions dominate in Y6 because cofacial $\pi$-orbital overlap between neighboring molecules gives rise to the strongest electronic couplings \cite{mahadevanAssessingIntraIntermolecular2024,gianniniRoleChargeTransfer2024,cerdáTuningExcitonDiffusion2025}.

\begin{figure}[ht]
\centering
\includegraphics[width=0.8\linewidth]{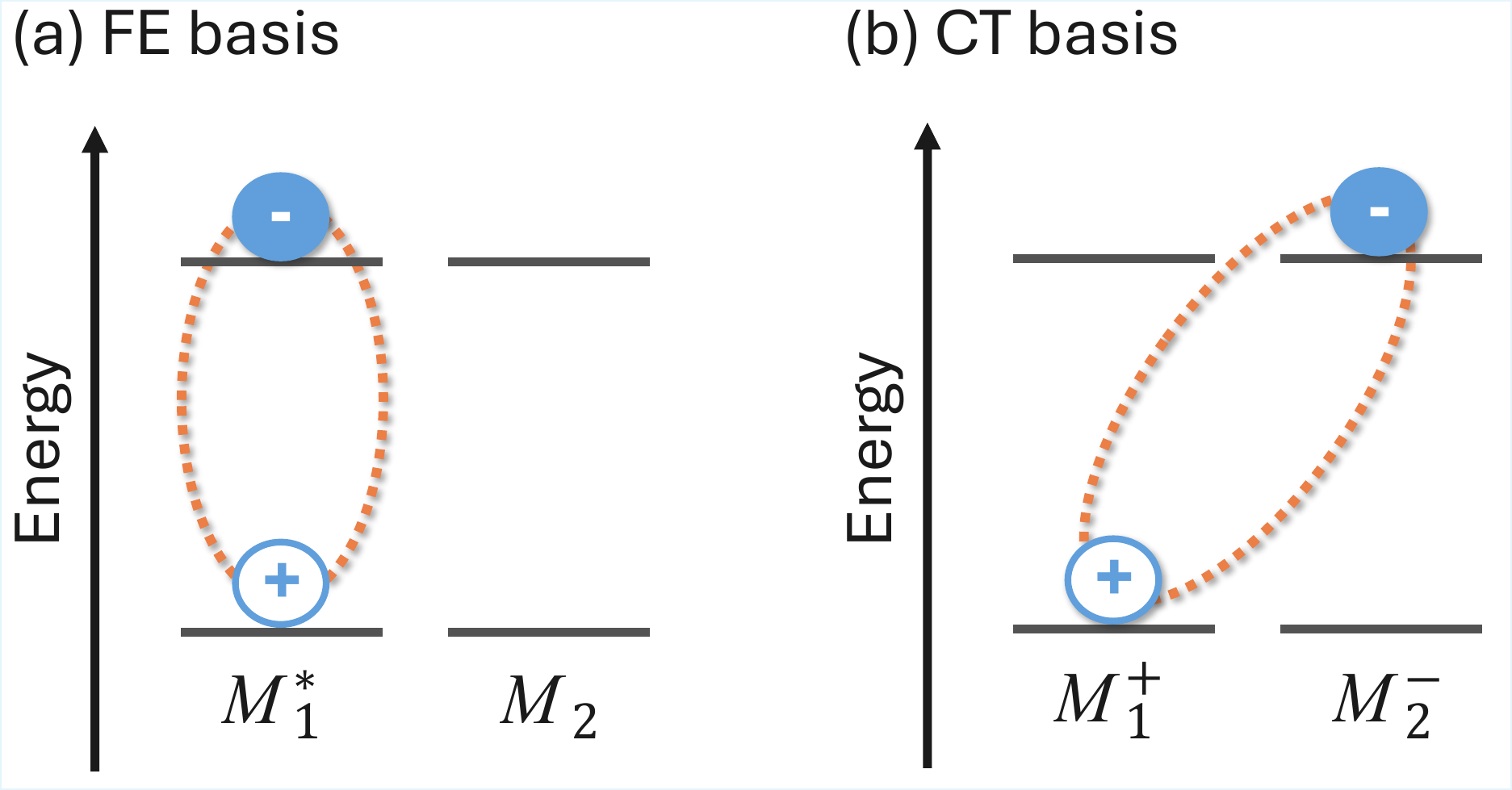}
\caption{
Schematic representation of the diabatic basis states for a molecular dimer.
(a) The electron and hole are bound on the same molecule in the FE basis configuration, where $M_1^*$ denotes a localized $\pi\text{--}\pi^*$ excitation.
(b) The electron and hole are separated across neighboring molecules ($M_1^+M_2^-$) in the CT basis configuration. The spatial separation of the electron and hole in the CT configuration reduces their Coulombic attraction, resulting in a higher excitation energy than the corresponding FE configuration.
}
\label{fig:configurations}
\end{figure}

The electronic Hamiltonian is written as
\begin{equation}
H_{\mathrm{el}} = H_{\mathrm{FE}} + H_{\mathrm{CT}} + H_{\mathrm{mix}},
\label{eq:Hel_partition}
\end{equation}
where the FE block is
\begin{equation}
H_{\mathrm{FE}}
=
\sum_n E_n^{\mathrm{FE}} |n\rangle\langle n|
+
\sum_{n\neq m} J_{nm}^{\mathrm{FE}}\,|n\rangle\langle m|,
\label{eq:H_FE}
\end{equation}
the CT block is
\begin{equation}
H_{\mathrm{CT}}
=
\sum_{n<m} E_{nm}^{\mathrm{CT}} |nm\rangle\langle nm|
+
\sum_{n<m}\sum_{u<\ell}
J_{nm,u\ell}^{\mathrm{CT}}\,|nm\rangle\langle u\ell|,
\label{eq:H_CT}
\end{equation}
and the FE and CT basis configurations are coupled through off-diagonal terms
\begin{equation}
H_{\mathrm{mix}}
=
\sum_{n,m}
\left(
V_{n,nm}\,|n\rangle\langle nm|
+
V_{nm,n}\,|nm\rangle\langle n|
\right).
\label{eq:H_mix}
\end{equation}
In these expressions, $E_n^{\mathrm{FE}}$ denote excitation energies localized on individual molecules, $E_{nm}^{\mathrm{CT}}$ is the energy of a charge-separated configuration, $J_{nm}^{\mathrm{FE}}$ is the intermolecular transition dipole coupling, $J_{nm,u\ell}^{\mathrm{CT}}$ is the coupling between CT configurations, and $V_{n,nm}$ represents electron- or hole-transfer matrix elements that interconvert neutral and charge-separated configurations. The CT basis states are defined for unordered molecular pairs $(n,m)$ with $n<m$, so that $|nm\rangle \equiv |mn\rangle$, and the corresponding energies satisfy $E_{nm}^{\mathrm{CT}} = E_{mn}^{\mathrm{CT}}$. The distinction between the FE and CT basis configurations is illustrated for a molecular dimer in Fig.~\ref{fig:configurations}.

The eigenstates of the overall electronic Hamiltonian can be expanded as a superposition of FE and CT configurations,
\begin{equation}
|\alpha\rangle
=
\sum_n C_n^{(\alpha)} |n\rangle
+
\sum_{n<m} D_{nm}^{(\alpha)} |nm\rangle .
\label{eq:eigen_expansion}
\end{equation}
The expansion coefficients define the FE and CT composition of each eigenstate. The FE fraction is expressed as
\begin{equation}
f_\alpha^{\mathrm{FE}}
=
\sum_n |C_n^{(\alpha)}|^2,
\label{eq:f_FE}
\end{equation}
and the CT fraction is
\begin{equation}
f_\alpha^{\mathrm{CT}}
=
\sum_{n<m} |D_{nm}^{(\alpha)}|^2.
\label{eq:f_CT}
\end{equation}
These quantities provide a convenient measure of the degree of FE--CT hybridization for each eigenstate.

\subsection{Linear Absorption Spectrum from the Mixed FE--CT Hamiltonian}
\label{sec:absorption_model}

We next construct an effective model for the linear absorption spectrum based on the mixed FE--CT Hamiltonian introduced above. While the electronic structure of molecular films can be described using models that explicitly include molecular degrees of freedom, a reduced FE--CT description provides a practical framework for extracting microscopic parameters from experimental absorption spectra. Within this representation, the Hamiltonian is written in $k$-space as the $2\times2$ matrix,
\begin{equation}
H(k)=
\begin{pmatrix}
E_{\mathrm{FE}} + 2J_{\mathrm{FE}}\cos(k) & V \\
V & E_{\mathrm{CT}} + 2J_{\mathrm{CT}}\cos(k)
\end{pmatrix},
\label{eq:Hk}
\end{equation}
where $E_{\mathrm{FE}}$ and $E_{\mathrm{CT}}$ are the FE and CT basis energies, $J_{\mathrm{FE}}$ and $J_{\mathrm{CT}}$ are the corresponding nearest-neighbor couplings, and $V$ is the FE--CT mixing interaction. The diagonal terms describe uncoupled FE and CT bands with the familiar cosine dispersion of one-dimensional nearest-neighbor tight-binding models, while the off-diagonal interaction $V$ hybridizes the two bands. Here, $k$ denotes the dimensionless wavevector obtained by scaling the physical wavevector by the intermolecular spacing. The first Brillouin zone extends from $-\pi$ to $\pi$, with $k=\pm\pi$ corresponding to an exciton wavefunction whose phase alternates by $\pi$ between neighboring molecules.

In this way, the parameters defined in Sec.~\ref{sec:fect_hamiltonian} are incorporated into a band representation of the FE--CT manifold. Although the assumption of one-dimensional periodicity neglects the directional dependence of the three-dimensional Y6 morphology and can overestimate the spatial extent of the excitons, the broad absorption linewidths at ambient conditions render the calculated spectra relatively insensitive to these approximations. In addition, both the FE and CT couplings, $J_{\mathrm{FE}}$ and $J_{\mathrm{CT}}$, are expected to be largest along the cofacial stacking direction due to the shorter intermolecular distances. By contrast, describing the transport mechanism requires an accurate estimate of the exciton size, as this distinguishes diffusive motion from delocalization-mediated exciton migration. These limitations are addressed below by introducing a method that interpolates between localized (diffusive) and delocalized (band-like) limits, consistent with transient localization arising from dynamic intermolecular disorder \cite{troisiChargeTransportRegimeCrystalline2006,ciuchiTransientLocalizationCrystalline2011}.

Diagonalization of Eq.~\eqref{eq:Hk} yields two hybridized exciton branches with energies $E_\alpha(k)$ and eigenvectors that determine the FE and CT character of each state. The transition dipole of eigenstate $\alpha$ arises from its FE component and is given by
\begin{equation}
\mu_\alpha(k)
=
\sum_n C_n^{(\alpha)}(k)\,\mu_n,
\label{eq:mu_alpha_k}
\end{equation}
where $\mu_n$ is the transition dipole associated with a localized FE state on site $n$. This expression reflects an intensity borrowing mechanism in which excitons with significant CT character acquire oscillator strength through mixing with FE states. In the slow-modulation (inhomogeneous broadening) limit \cite{mukamelPrinciplesNonlinearOptical1995}, each transition is described by a Gaussian lineshape, and the absorption spectrum is written as
\begin{equation}
\begin{split}
A(\omega)
&=
A_0
\sum_{\alpha}
\int dk\,
\left|
\mu_\alpha(k)
\right|^2
\frac{1}{\sqrt{2\pi}\,\Delta_\alpha(k)}
\\
&\qquad\times
\exp\!\left[
-\frac{(\hbar\omega - E_\alpha(k))^2}
{2\Delta_\alpha^2(k)}
\right],
\end{split}
\label{eq:abs_gaussian_k}
\end{equation}
where $A_0$ is an overall amplitude and the integral over $k$ effectively accounts for the distribution of excitonic states along the dominant stacking direction.

The effects of environmental fluctuations on the absorption lineshape are characterized by the reorganization energy. Within the classical high-temperature limit, the Gaussian broadening parameter in Eq.~\eqref{eq:abs_gaussian_k} is related to the reorganization energy through
\begin{equation}
\lambda_\alpha(k) = \frac{\Delta_\alpha(k)^2}{2k_B T}.
\label{eq:lambda}
\end{equation}
The variance of the exciton transition energy is expressed as a weighted average of the FE and CT contributions,
\begin{equation}
\Delta^2_\alpha(k)
=
2k_B T
\left[
f_\alpha^{\mathrm{FE}}(k)\,\lambda_{\mathrm{FE}}
+
f_\alpha^{\mathrm{CT}}(k)\,\lambda_{\mathrm{CT}}
\right],
\label{eq:Delta_alpha}
\end{equation}
where $\lambda_{\mathrm{FE}}$ and $\lambda_{\mathrm{CT}}$ are the reorganization energies of the FE and CT configurations.

Equation~\eqref{eq:abs_gaussian_k} provides a practical expression for fitting experimental absorption spectra and extracting microscopic electronic structure parameters. This reciprocal-space formulation captures the same excitonic bands as a large real-space tight-binding model of an extended molecular stack while reducing the computational cost from repeated diagonalization of a large Hamiltonian to evaluation of a $2\times2$ Hamiltonian over the Brillouin zone. The adjustable parameters are the effective energies $E_{\mathrm{FE}}$ and $E_{\mathrm{CT}}$, the nearest-neighbor couplings $J_{\mathrm{FE}}$ and $J_{\mathrm{CT}}$, the mixing interaction $V$, and the reorganization energies $\lambda_{\mathrm{FE}}$ and $\lambda_{\mathrm{CT}}$. The same parameters also determine the Redfield population transfer rates, ensuring consistency between the spectral and dynamical descriptions.

\subsection{Redfield Population Transfer and Turnover}
\label{sec:redfield}

Strong intermolecular coupling in non-fullerene acceptors such as Y6 promotes delocalization of excitonic states through coherent mixing of local excitations; however, coupling to environmental fluctuations, particularly for CT configurations with larger reorganization energies, tends to localize excitons onto a small number of molecules. Redfield theory provides a microscopic description of population transfer between delocalized exciton states \cite{BreuerPetruccione2002,yangInfluencePhononsExciton2002,MayKuhn2011,balzerMechanismDelocalizationEnhancedExciton2023,abramaviciusDynamicsPhotoexcitedExcitonphonon2026}. The present model extends this description by incorporating transient localization, thereby capturing the competition between electronic delocalization and environmental fluctuations that gives rise to a turnover in the exciton transfer rate.

In the Redfield framework, the transition rate from exciton $|\beta\rangle$ to $|\alpha\rangle$ is given by
\begin{equation}
R_{\alpha\beta}
=
\frac{1}{\hbar^2}
\sum_{\eta}
|a_\eta^{(\alpha)}|^2 |a_\eta^{(\beta)}|^2
S_{\eta}(\omega_{\alpha\beta}),
\label{eq:redfield_general_main}
\end{equation}
where $\eta$ indexes both FE and CT diabatic configurations,
$a_\eta^{(\alpha)}$ denotes the amplitude of configuration $\eta$ in eigenstate $|\alpha\rangle$, and $\omega_{\alpha\beta}=(E_\alpha-E_\beta)/\hbar$. Assuming Debye relaxation of the bath, the corresponding spectral density is
\begin{equation}
S_\eta(\omega)=
\frac{4\lambda_\eta k_B T\,\Lambda}
{\omega^2+\Lambda^2},
\label{eq:S_lorentz_lambda}
\end{equation}
where $\Lambda$ is the bath relaxation rate and $\lambda_\eta$ is the reorganization energy associated with diabatic configuration $\eta$. Equation~\eqref{eq:S_lorentz_lambda} shows that the reorganization energy $\lambda_\eta$ governs the magnitude of the system--bath coupling, whereas the relaxation rate $\Lambda$ determines the spectral range over which environmental fluctuations efficiently drive exciton transitions.

Separating the contributions from FE and CT configurations yields
\begin{equation}
\begin{split}
R_{\alpha\beta}
&=
\frac{S_{\mathrm{FE}}(\omega_{\alpha\beta})}{\hbar^2}
\sum_n
|C_n^{(\alpha)}|^2
|C_n^{(\beta)}|^2
\\
&\quad+
\frac{S_{\mathrm{CT}}(\omega_{\alpha\beta})}{\hbar^2}
\sum_{n<m}
|D_{nm}^{(\alpha)}|^2
|D_{nm}^{(\beta)}|^2,
\end{split}
\label{eq:redfield_decomposition_main}
\end{equation}
where $S_{\mathrm{FE}}$ and $S_{\mathrm{CT}}$ are the spectral densities associated with the two classes of diabatic configurations. To simplify the rate expression, we introduce the average CT fraction of the two eigenstates involved in the transition,
\begin{equation}
\bar{f}_{\mathrm{CT}}=\frac{f_\alpha^{\mathrm{CT}}+f_\beta^{\mathrm{CT}}}{2},
\end{equation}
where $f_\alpha^{\mathrm{CT}}$ is defined in Eq.~\eqref{eq:f_CT}. Assuming that the spectral density varies smoothly with CT character and that transitions occur predominantly between states with similar CT composition, the rate is approximated by the factorized form,
\begin{equation}
R_{\alpha\beta}(\bar{f}_{\mathrm{CT}})
\approx
\frac{1}{\hbar^2}S_{\mathrm{eff}}(\omega_{\alpha\beta};\bar{f}_{\mathrm{CT}})\,
\mathcal O_{\alpha\beta}(\bar{f}_{\mathrm{CT}}),
\label{eq:k_factorized_main}
\end{equation}
where
\begin{equation}
S_{\mathrm{eff}}(\omega;\bar{f}_{\mathrm{CT}})
=
(1-\bar{f}_{\mathrm{CT}})S_{\mathrm{FE}}(\omega)
+\bar{f}_{\mathrm{CT}} S_{\mathrm{CT}}(\omega)
\end{equation}
is an effective system--bath coupling, and $\mathcal O_{\alpha\beta}$ is an overlap factor that depends on the degree to which the two eigenstates share the same diabatic configurations (see Appendix~\ref{app:overlap}). The overlap factor $\mathcal O_{\alpha\beta}$ is next parameterized in terms of the average CT character.

The spatial extent of exciton eigenstates is governed primarily by the strength of the system--bath coupling, which increases with the CT character through its larger reorganization energy. Accordingly, the overlap factor $\mathcal O_{\alpha\beta}$ is approximated as a function of $\bar{f}_{\mathrm{CT}}$. Using Eq.~\eqref{eq:lambda}, the effective variance of the disorder can be written as
\begin{equation}
\sigma_{\mathrm{eff}}^2(\bar{f}_{\mathrm{CT}})
=
2k_BT\left[(1-\bar{f}_{\mathrm{CT}})\lambda_{\mathrm{FE}}+\bar{f}_{\mathrm{CT}}\lambda_{\mathrm{CT}}\right].
\label{eq:sigmaeff_main}
\end{equation}
The influence of disorder on the spatial extent of the exciton is characterized by an effective delocalization size that interpolates between the weak- and strong-disorder limits,
\begin{equation}
N_{\mathrm{eff}}(\bar{f}_{\mathrm{CT}})
=
N_{\min}
+
\frac{N_{\max}-N_{\min}}
{1+\left[\sigma_{\mathrm{eff}}(\bar{f}_{\mathrm{CT}})/\sigma_c\right]^2},
\label{eq:Neff_main}
\end{equation}
where $N_{\max}$ is the delocalized limit, $N_{\min}$ is the localized limit, and the phenomenological parameter $\sigma_c$ sets the crossover scale. As shown in Appendix~\ref{app:localization}, the overlap factor is approximated by
\begin{equation}
\mathcal O(\bar{f}_{\mathrm{CT}})
\approx
\frac{
P(\bar{f}_{\mathrm{CT}})
}
{
N_{\mathrm{eff}}(\bar{f}_{\mathrm{CT}})
}.
\end{equation}
Substituting this expression into Eq.~\eqref{eq:k_factorized_main} yields
\begin{equation}
R_{\mathrm{ET}}(\bar{f}_{\mathrm{CT}})
\approx
\frac{
\left[(1-\bar{f}_{\mathrm{CT}})S_{\mathrm{FE}}(\omega)
+\bar{f}_{\mathrm{CT}}S_{\mathrm{CT}}(\omega)\right]
P(\bar{f}_{\mathrm{CT}})
}
{\hbar^2N_{\mathrm{eff}}(\bar{f}_{\mathrm{CT}})}.
\label{eq:k_final_main}
\end{equation}
Equation~\eqref{eq:k_final_main} reveals the origin of the turnover in the exciton transfer rate with increasing CT character: the spectral density increases approximately linearly with $\bar{f}_{\mathrm{CT}}$, whereas the spatial overlap factor $P(\bar{f}_{\mathrm{CT}})$ decreases exponentially due to disorder-induced localization.

\section{Experimental Results and Discussion}

\subsection{Parameterization of the Mixed FE--CT Hamiltonian Using the Absorption Spectrum}
\label{sec:abs_fit}

The model developed in Sec.~\ref{sec:absorption_model} provides a framework for extracting physically meaningful empirical parameters from the absorption spectrum. In particular, we seek to quantify the CT fraction $f_\alpha^{\mathrm{CT}}(k)$, which will later be connected to the non-diffusive character of the Redfield population transfer rates. Because the absorption lineshape reflects both the distribution of eigenstate energies and their oscillator strengths, it provides direct experimental access to the degree of FE--CT mixing. In addition, the linewidth of each transition reflects its coupling to environmental fluctuations through the reorganization energy, which generally increases with CT character and plays a central role in the resulting relaxation dynamics.

While the CT bandwidth is only weakly constrained by the absorption spectrum because of the low oscillator strength of CT states, the electron and hole transfer integrals in Y6 have been computed for a distribution of molecular dimers in prior work \cite{gianniniRoleChargeTransfer2024}. Due to the strong distance dependence of these couplings, CT contributions are dominated by nearest neighbors with cofacial orientations. Therefore, we adopt representative values of the electron and hole transfer integrals, $t_e \approx 60~\mathrm{meV}$ and $t_h \approx 40~\mathrm{meV}$, and define the effective CT coupling as $J_{\mathrm{CT}} \approx \sqrt{t_e t_h}$. This yields $J_{\mathrm{CT}} \approx 50~\mathrm{meV}$, consistent with the weak dispersion of CT states and their localized character. The remaining model parameters are then obtained by fitting the calculated absorption spectrum to experiment.

\begin{figure*}[ht]
\centering
\includegraphics[width=0.75\linewidth]{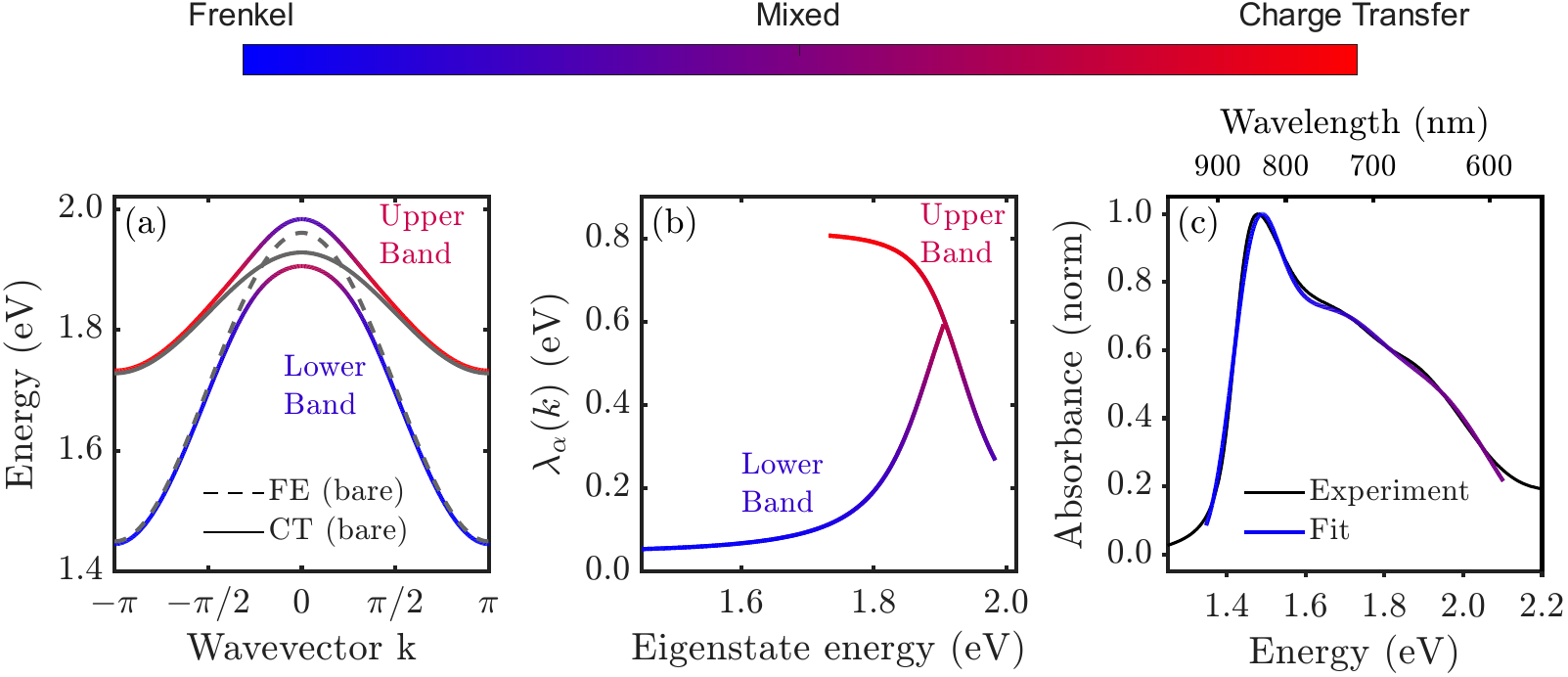}
\caption{
(a) Electronic band structure calculated from the mixed FE--CT Hamiltonian in Eq.~\eqref{eq:Hk} as a function of the dimensionless wavevector $k$. The dashed and solid black curves denote the bare FE and CT bands, respectively, while the colored curves show the hybridized eigenstates. The color scale indicates the CT fraction $f_\alpha^{\mathrm{CT}}(k)$, ranging from FE-like (blue) to CT-like (red). 
(b) Reorganization energy $\lambda_\alpha(k)$ of each eigenstate as a function of eigenstate energy. The reorganization energy follows the charge--transfer character of the state, increasing with $f_\alpha^{\mathrm{CT}}(k)$ according to Eq.~\eqref{eq:Delta_alpha}. 
(c) Linear absorption spectrum of the Y6 film together with the fit obtained from the FE--CT model. Although CT configurations contribute little direct oscillator strength, they influence the spectral lineshape through intensity borrowing and enhanced environmental coupling, leading to energy-dependent broadening.
}
\label{absorbance}
\end{figure*}

The electronic structure and associated fit are illustrated in Fig.~\ref{absorbance}, with the model parameters summarized in Table~\ref{tab:FECT_fit_parameters}. As shown in Fig.~\ref{absorbance}(a), the underlying diabatic FE energy is lower than the CT energy by $\sim$0.12~eV, reflecting the greater electron--hole binding energy of the $\pi$--$\pi^*$ excitation (see Fig.~\ref{fig:configurations}). The bare FE band also exhibits a larger dispersion than the CT band, consistent with the large intermolecular coupling $J_{\mathrm{FE}}$ within the FE block of the Hamiltonian \cite{zhangDelocalizationExcitonElectron2020,westbrookSolidStatePackingControls2025} and the small CT bandwidth inferred from literature values of intermolecular transfer integrals \cite{gianniniRoleChargeTransfer2024}. Hybridization between these manifolds is generally weak but becomes appreciable near the avoided crossing, where the eigenstates acquire mixed FE--CT character, as reflected by the smooth variation of the color scale. The fitted FE--CT coupling of $V \sim 0.035~\mathrm{eV}$ indicates moderate mixing, sufficient to redistribute the excitonic character near resonance while preserving predominantly FE- and CT-like states away from the intersection of the diabatic surfaces.

Although the band structure is one-dimensional, the behavior near the avoided crossing reflects the hybridization between FE and CT configurations, which follows directly from their finite coupling. While the detailed distribution of mixed states depends on dimensionality, the presence of hybridization itself is not an artifact of the reduced model. The present approach focuses on the cofacial stacking direction, which provides the dominant contribution to intermolecular coupling in the film. Both the fitted FE coupling $J_{\mathrm{FE}}$ and the assumed CT coupling $J_{\mathrm{CT}} \sim 50~\mathrm{meV}$ are consistent with the molecular packing geometry of Y6 \cite{gianniniRoleChargeTransfer2024}.

The reorganization energy associated with each eigenstate is shown in Fig.~\ref{absorbance}(b). The FE and CT diabatic configurations are characterized by distinct couplings to environmental fluctuations, with $\lambda_{\mathrm{CT}}$ more than an order of magnitude larger than $\lambda_{\mathrm{FE}}$, as determined from the absorption fit (Table~\ref{tab:FECT_fit_parameters}). Because the eigenstates are hybridized, the effective reorganization energy $\lambda_\alpha(k)$ follows the CT fraction $f_\alpha^{\mathrm{CT}}(k)$ according to Eq.~\eqref{eq:Delta_alpha}. Consequently, the lower and upper exciton bands exhibit increased and decreased reorganization energies, respectively, in the vicinity of the avoided crossing between the diabatic surfaces.

Having established the electronic structure and reorganization energies, the resulting absorption spectrum is shown in Fig.~\ref{absorbance}(c). The oscillator strength is dominated by the FE component of each eigenstate, as indicated by Eq.~\eqref{eq:mu_alpha_k}, leading to a spectrum that is composed primarily of FE-like (blue) states. Although CT configurations contribute little direct oscillator strength, their contribution to the eigenstates increases the effective reorganization energy and broadens the spectral lineshape. Intensity borrowing is most evident on the high-energy side of the spectrum, where states with appreciable CT character acquire oscillator strength through hybridization with FE configurations. FE--CT hybridization plays a crucial indirect role by suppressing the exchange narrowing characteristic of purely FE molecular aggregates, thereby enhancing the absorption linewidth. This broadening is essential for reproducing the smooth, asymmetric absorption profile observed experimentally. Furthermore, the fit indicates that the dominant influence of CT states on the absorption spectrum is hybridization-induced broadening, rather than superradiant H- and J-type aggregation.

\begin{table}[t]
\centering
\caption{Fitting parameters for the Y6 absorption spectrum}
\begin{tabular}{lc}
\hline
\textsuperscript{a}Parameter & Value \\
\hline
$E_{\mathrm{FE}}$ (eV) & $1.705 \pm 0.001$ \\
$E_{\mathrm{CT}}$ (eV) & $1.828 \pm 0.004$ \\
$J_{\mathrm{FE}}$ (eV) & $0.128 \pm 0.001$ \\
$V$ (eV) & $0.0353 \pm 0.004$ \\
$\lambda_{\mathrm{FE}}$ (eV) & $0.042 \pm 0.002$ \\
$\lambda_{\mathrm{CT}}$ (eV) & $0.819 \pm 0.039$ \\
\hline
\end{tabular}
\parbox{0.9\linewidth}{\footnotesize
\textit{\textsuperscript{a}} Uncertainties denote one standard deviation estimated from the fit covariance matrix. 
}
\label{tab:FECT_fit_parameters}
\end{table}

\subsection{Transient Absorption Results}
\label{sec:taresults}

Transient absorption spectroscopy is used to characterize the dynamics of photoexcited states and their evolution in the presence of an interfacial hole-transport layer. We first present broadband spectra to establish the spectral signatures of ground-state bleach and excited-state absorption. The probe window is then restricted to a narrower spectral region, enabling a quantitative analysis of interfacial charge-transfer dynamics through quenching measurements.

\begin{figure}[ht]
\centering
\includegraphics[width=1.0\linewidth]{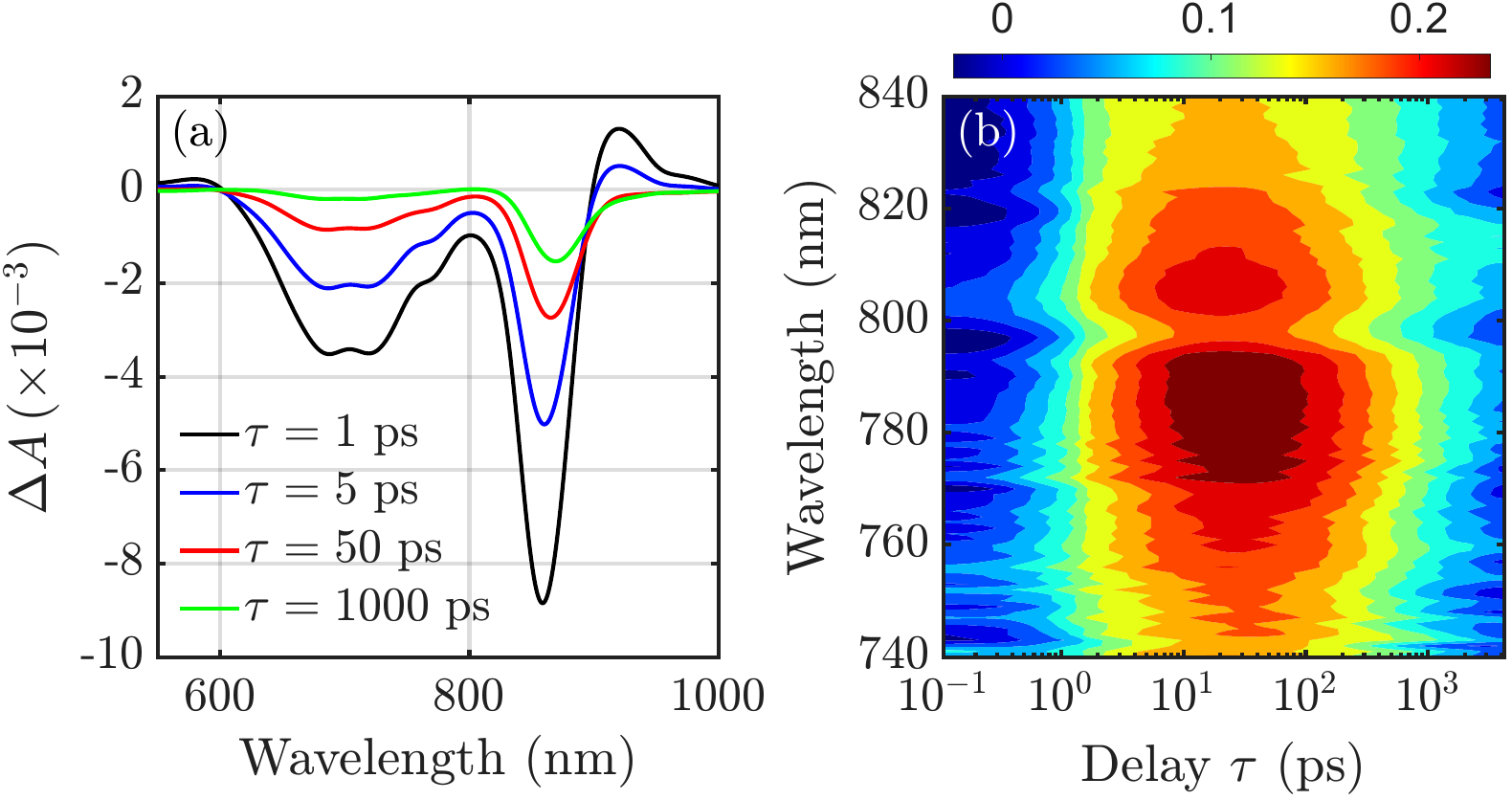}
\caption{
(a) Transient absorption spectra of Y6 films at selected pump--probe delay times. The spectra exhibit a negative signal at shorter wavelengths, attributed primarily to ground-state bleach, and a photoinduced absorption feature at longer wavelengths. A minimum in the signal magnitude near 800~nm arises from partial cancellation between these contributions. (b) Difference signal $Q(\lambda,\tau)$ obtained from normalized transient absorption measurements acquired with and without the CuSCN hole-transport layer, highlighting the contribution from interfacial charge separation.
}
\label{fig:taspectra}
\end{figure}

The transient absorption spectra of a neat Y6 film shown in Fig.~\ref{fig:taspectra}(a) exhibit a negative signal throughout the 600--900~nm spectral range, with excited-state absorption beyond 900~nm. The negative signal is attributed to ground-state bleach rather than stimulated emission because rapid excited-state relaxation shifts the stimulated emission to the spectral region corresponding to the steady-state emission, which lies predominantly beyond the probe window at wavelengths exceeding 900~nm \cite{zouInsightExcitationStates2020,xieWhyDoesY62024,kumarMorphologicalControlY62025}. Notably, the spectral lineshape is not simply an inverted replica of the steady-state absorption spectrum over the 600--900~nm range, indicating interference among multiple nonlinear optical contributions. In particular, the decrease in signal magnitude near $\sim 800$~nm is consistent with partial cancellation of the ground-state bleach by a photoinduced absorption band in this spectral region. A similar minimum in the transient absorption signal has been reported previously for Y6 films, where it was attributed to a polaronic resonance arising from the evolution of excited states with mixed FE--CT character \cite{gianniniRoleChargeTransfer2024}.

Quenching measurements are conducted by comparing the temporal profiles of transient absorption signals acquired for Y6 films in the absence and presence of a CuSCN hole-transport layer. Fig.~\ref{fig:taspectra}(b) presents spectra obtained using a separate experimental configuration with a narrower probe bandwidth. The plotted quantity is the difference between normalized signals,
\begin{equation}
Q(\lambda,\tau) = \Delta \tilde{A}_{\mathrm{Y6:HTL}}(\lambda,\tau) - \Delta \tilde{A}_{\mathrm{Y6}}(\lambda,\tau),
\label{eq:quenchingQ}
\end{equation}
where $\Delta \tilde{A}(\lambda,\tau) = \Delta A(\lambda,\tau)/\Delta A(\lambda,\tau_0)$ is obtained from the measured transient absorption response $\Delta A(\lambda,\tau)$ with $\tau_0 = 0.1$~ps. Although the overall signal lineshape is only weakly perturbed, the quenching effect is most pronounced in the 770--820~nm range, near the polaronic excited-state absorption resonance. Moreover, the sign of $Q(\lambda,\tau)$ suggests that the CuSCN layer slows the relaxation of the transient absorption response, consistent with the delayed equilibration of the electronic population illustrated in Fig.~\ref{fig:tascheme}.

\begin{figure*}[ht]
\centering
\includegraphics[width=0.9\linewidth]{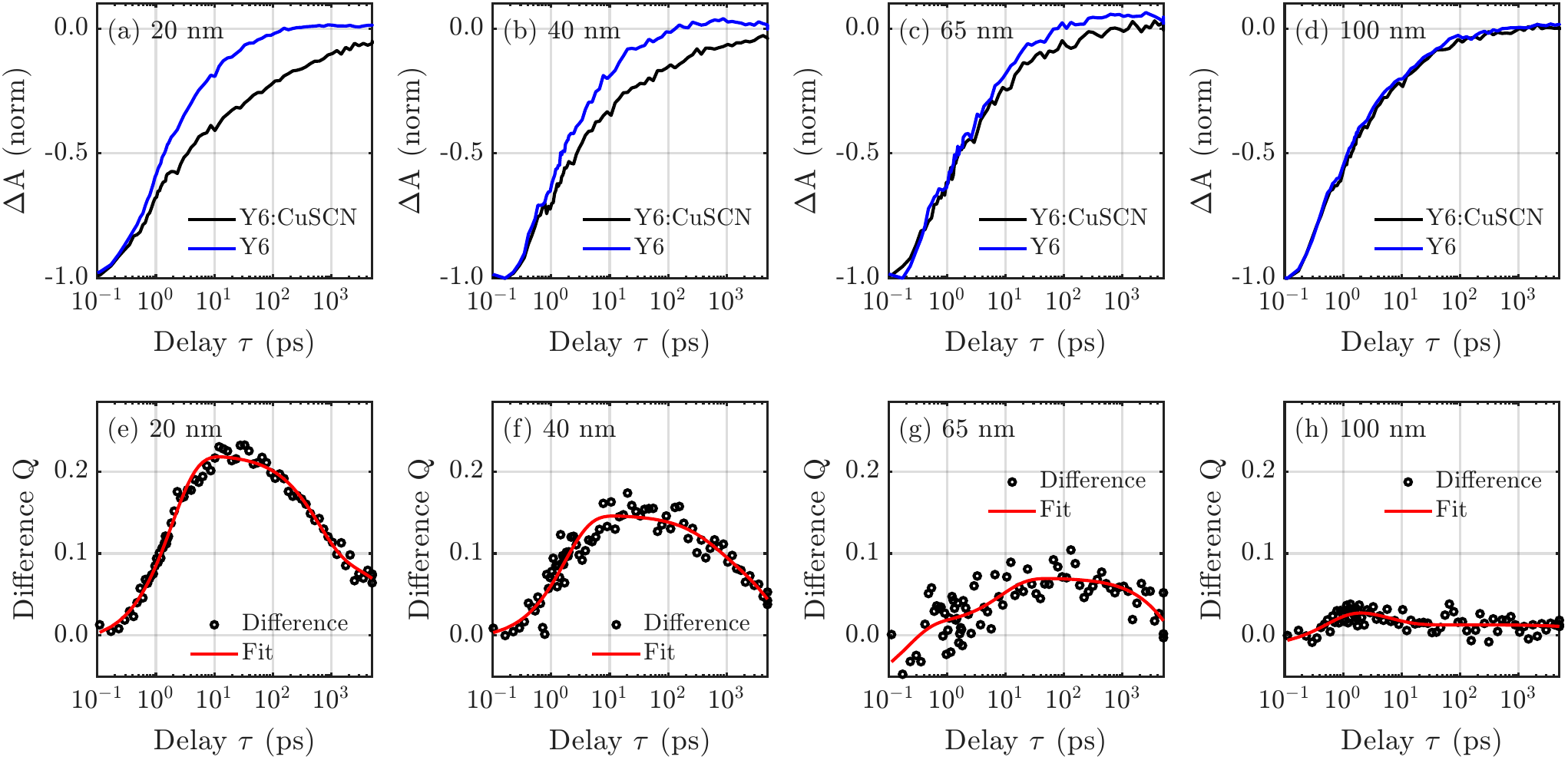}
\caption{
Transient absorption dynamics of Y6 films with varying thicknesses measured with and without CuSCN hole-transport layers.
(a)--(d) Normalized temporal profiles at a detection wavelength of 810~nm for the indicated film thicknesses.
(e)--(h) Quenching profiles, $Q$, together with fits to Eq.~\eqref{eq:rise_double_decay}.
The fitted curves are used to determine the characteristic hole-transfer time and quenching efficiency.
}
\label{quenching}
\end{figure*}

The quenching kinetics are quantified by analyzing the temporal profiles $Q(\lambda,\tau)$ at a detection wavelength of 810~nm, where significant quenching is observed with minimal scattering from the 795-nm pump pulse. The signals presented in Fig.~\ref{quenching} are fitted using
\begin{equation}
\begin{split}
F(t)
&=
B_0
+
\left(1-\exp\left(-t/t_r\right)\right)
\\
&\qquad\times
\left[
B_1\exp\left(-t/\tau_{d1}\right)
+
B_2\exp\left(-t/\tau_{d2}\right)
\right],
\end{split}
\label{eq:rise_double_decay}
\end{equation}
where $B_0$ accounts for a baseline offset, $t_r$ describes the initial rise associated with hole transfer, and $B_1$ and $B_2$, together with the associated time constants $\tau_{d1}$ and $\tau_{d2}$, characterize the decay of the charge-separated population through recombination. The model parameters were obtained by nonlinear least-squares fitting. Parameter uncertainties were estimated from the covariance matrix, $\mathbf{C} = s^2 (\mathbf{D}^\mathrm{T}\mathbf{D})^{-1}$, where $\mathbf{D}$ is the Jacobian and $s^2$ is the residual variance. The uncertainties reported in Table~\ref{tab:fit_parameters} correspond to one standard deviation obtained from the square roots of the diagonal elements of $\mathbf{C}$.

The difference between transient absorption signals measured in the presence and absence of the hole-transport layer isolates the contribution from interfacial charge separation. Since the signal magnitude is proportional to the excited-state population, the difference $Q(\lambda,\tau)$ provides a direct measure of the transient charge-separated population. The peak buildup time $t_{\mathrm{peak}}$ and peak amplitude $Q_{\max}$ were extracted from the fitted response $F(t)$, thereby reducing the influence of experimental noise. The peak time was determined numerically as the maximum of the fitted function evaluated on a dense time grid, and $Q_{\max}=F(t_{\mathrm{peak}})$. Uncertainties in these quantities were obtained by Monte Carlo propagation of the parameter covariance matrix, which accounts for correlations among the fitted parameters and the nonlinear dependence of the peak characteristics on the model. This approach provides a compact representation of the buildup and maximum population of charge-separated states, enabling direct comparison of hole-transfer dynamics across film thicknesses.

\begin{table*}[t]
\centering
\caption{Hole-transfer fitting parameters extracted for Y6 films of varying thicknesses}
\begin{tabular}{lcccc}
\hline
\textsuperscript{a}Parameter & 20 nm & 40 nm & 65 nm & 95 nm \\
\hline
$B_0$ & $-0.009 \pm 0.004$ & $-0.005 \pm 0.007$ & $-0.067 \pm 0.025$ & $-0.014 \pm 0.009$ \\
$t_r$ (ps) & $1.90 \pm 0.08$ & $1.74 \pm 0.18$ & $0.20 \pm 0.09$ & $0.59 \pm 0.28$ \\
$B_1$ & $0.126 \pm 0.004$ & $0.042 \pm 0.009$ & $-0.057 \pm 0.009$ & $0.021 \pm 0.009$ \\
$\tau_{d1}$ (ps) & $589 \pm 16$ & $551 \pm 30$ & $8.39 \pm 3.74$ & $6.63 \pm 4.52$ \\
$B_2$ & $0.104 \pm 0.006$ & $0.110 \pm 0.012$ & $0.137 \pm 0.027$ & $0.027 \pm 0.009$ \\
$\tau_{d2}$ (ns) & $17.64 \pm 0.02$ & $6.26 \pm 0.03$ & $10.45 \pm 0.04$ & $70.27 \pm 0.02$ \\
$t_{\mathrm{peak}}$ (ps) & $12.0 \pm 0.4$ & $11.9 \pm 0.9$ & $52.5 \pm 18.2$ & $2.0 \pm 5.9$ \\
$Q_{\max}$ & $0.218 \pm 0.002$ & $0.146 \pm 0.004$ & $0.070 \pm 0.004$ & $0.027 \pm 0.004$ \\
\hline
\end{tabular}

\vspace{1mm}
\parbox{0.8\linewidth}{\footnotesize
\textit{\textsuperscript{a}} Uncertainties denote one standard deviation from the fit.
}
\label{tab:fit_parameters}
\end{table*}

To quantify the spatial extent over which excitations contribute to interfacial charge transfer, we estimate an effective capture thickness $\ell_{\mathrm{cap}}$ from the measured quenching efficiency $Q_{\max}$. Accounting for the finite optical penetration depth $L_{\mathrm{opt}}\sim90$~nm, the relationship between $Q_{\max}$ and $\ell_{\mathrm{cap}}$ is given by
\begin{equation}
\ell_{\mathrm{cap}} = L_{\mathrm{opt}}\, Q_{\max}(d)\left[1 - \exp\left(-d/L_{\mathrm{opt}}\right)\right],
\label{eq:quench_depth}
\end{equation}
where $d$ is the film thickness, and we have assumed $\ell_{\mathrm{cap}} \ll L_{\mathrm{opt}}$. Using the experimental values of $Q_{\max}$ for films exhibiting quenching efficiencies greater than 5\%, we obtain an average capture thickness $\ell_{\mathrm{cap}} = 4.03 \pm 0.74$ nm. This result indicates that the hole transfer process is dominated by excitations generated within a few nanometers of the Y6:CuSCN interface. The weak dependence of the quenching time $t_{\mathrm{peak}}$ on film thickness is inconsistent with purely diffusive transport, which would predict characteristic arrival times that scale as $d^2$. Instead, the data support an ultrafast transport mechanism in which the thickness dependence of $Q_{\max}$ reflects the fraction of the initial excitation profile lying within this finite capture region.

\subsection{Turnover-Limited Exciton Transport and Interfacial Quenching}
\label{sec:turnover_quenching}

The FE--CT relaxation mechanism is connected to the experimentally observed quenching dynamics through a minimal kinetic model describing exciton migration to the CuSCN interface followed by hole transfer. We first use the parameters extracted from the absorption spectrum to evaluate the reduced Redfield rate in Eq.~\eqref{eq:k_final_main}, thereby establishing the predicted turnover as a function of CT character. The resulting transfer rate is then incorporated into a kinetic model describing the transient absorption quenching dynamics.

The Redfield exciton transfer rate $R_{\mathrm{ET}}$ is parameterized using experimentally constrained and physically motivated quantities. The electronic energies, intermolecular couplings, and reorganization energies are taken directly from the fit of the absorption spectrum (Table~\ref{tab:FECT_fit_parameters}). Additional parameters governing the localization and spatial overlap of excitons are chosen based on physically reasonable estimates. In particular, the crossover between delocalized and localized behavior is controlled by a characteristic disorder scale $\sigma_c$ [Eq.~\eqref{eq:Neff_main}], which we take as the mean of the FE and CT coupling strengths, $\sigma_c = (J_{\mathrm{FE}} + J_{\mathrm{CT}})/2 \approx 0.089~\mathrm{eV}$. This choice reflects the role of FE--CT hybridization in the onset of localization. The exciton participation size $N_{\mathrm{eff}}(\sigma)$ represents the effective number of FE--CT diabatic configurations contributing to the eigenvector. The upper bound of $N_{\max} = 28$ is motivated by estimates of transient exciton delocalization over $\sim 10$ molecular units in Y6 \cite{gianniniExcitonTransportMolecular2022} and related molecular aggregates \cite{scholesLimitsExcitonDelocalization2019}, together with the inclusion of nearest-neighbor CT configurations, which expand the Hilbert space beyond the number of molecular sites. The lower bound $N_{\min} \sim 1$--2 represents localization to a small number of configurations in the strong-disorder limit. The characteristic separation $r_0$ entering Eq.~\eqref{eq:P_fct} is taken to be one nearest-neighbor intermolecular spacing, consistent with the dominant role of cofacially stacked molecular pairs in exciton transport. Although these parameters are not independently fitted, the resulting turnover in $R_{\mathrm{ET}}$ is robust against moderate variations.

The spectral density entering the Redfield rate [Eq.~\eqref{eq:S_lorentz_lambda}] is evaluated at a representative transition frequency $\omega=\Delta E/\hbar$, where $\Delta E$ is the characteristic exciton energy spacing and $\Lambda$ is the bath relaxation rate. In this form, the frequency-dependent denominator depends on the dimensionless combination $\Delta E/(\hbar\Lambda)$, which determines the extent to which environmental fluctuations can drive population transfer between non-degenerate exciton states. Because the model does not explicitly resolve the full distribution of exciton energy spacings or bath dynamics, these quantities are treated as effective parameters constrained by physical considerations. Given intermolecular couplings of order $J_{\mathrm{FE}}$ and $J_{\mathrm{CT}}$, and an exciton manifold comprising a finite number of hybrid FE--CT configurations, the resulting energy gaps are on the order of tens of meV. Therefore, we take $\Delta E = 50~\mathrm{meV}$ as the characteristic spacing between pairs of exciton states involved in the non-radiative transitions. The correlation time $\Lambda^{-1} = 100$ fs is consistent with sub-picosecond electronic energy fluctuations obtained from simulations of electron--phonon coupling in Y6 \cite{gianniniExcitonTransportMolecular2022}. Importantly, moderate variations in $\Delta E$ and $\Lambda$ primarily rescale the overall magnitude of $R_{\mathrm{ET}}$ without changing the turnover behavior predicted as a function of $\bar{f}_{\mathrm{CT}}$.

\begin{figure}[ht]
\centering
\includegraphics[width=1.0\linewidth]{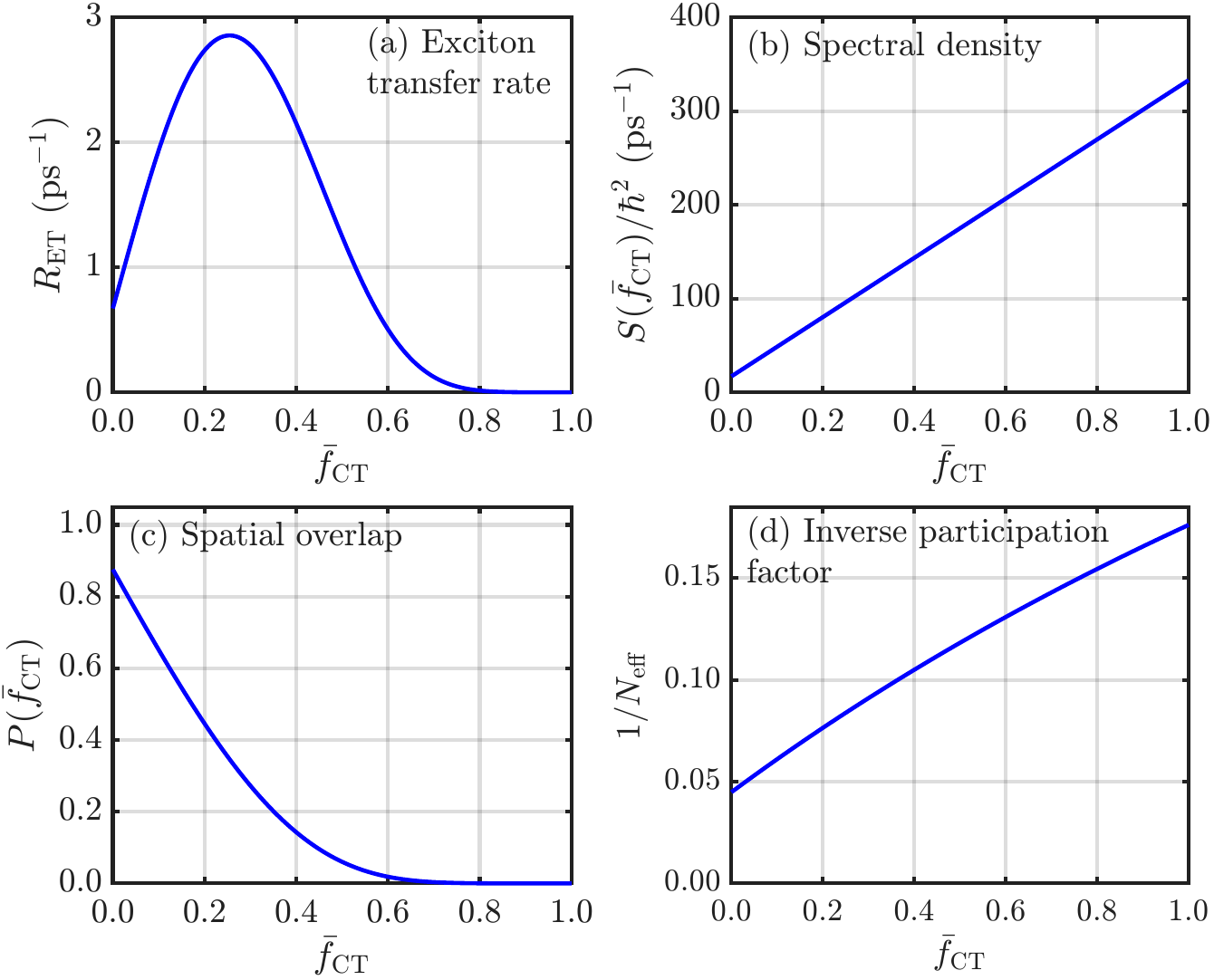}
\caption{(a) The exciton transfer rate, $R_{\mathrm{ET}}$, computed using Eq.~\eqref{eq:k_final_main} exhibits a pronounced turnover as a function of the average fraction of CT basis configurations, $\bar{f}_{\mathrm{CT}}$. (b) The bath spectral density, $S(\bar{f}_{\mathrm{CT}})$, increases with CT character because of enhanced system--environment coupling. (c) The spatial overlap factor, $P(\bar{f}_{\mathrm{CT}})$, decreases as disorder increases, indicating stronger exciton localization. (d) The inverse participation factor, $1/N_{\mathrm{eff}}$, likewise indicates reduced exciton delocalization with increasing CT character. The overall transfer rate results from the competition among these effects, showing that intermediate FE--CT hybridization maximizes the exciton transfer rate. }
\label{fig:turnover}
\end{figure}

The resulting dependence of the exciton transfer rate on the average CT fraction $\bar{f}_{\mathrm{CT}}$ is shown in Fig.~\ref{fig:turnover}. Following Eq.~\eqref{eq:k_final_main}, the overall rate in Fig.~\ref{fig:turnover}(a) is decomposed into contributions from the bath spectral density, spatial overlap, and inverse participation factor in Figs.~\ref{fig:turnover}(b)--\ref{fig:turnover}(d). These components reflect distinct physical effects. The spectral density increases with $\bar{f}_{\mathrm{CT}}$ due to stronger coupling of CT configurations to environmental fluctuations, thereby promoting non-radiative transitions through efficient energy exchange with the bath. Increasing CT character also enhances energetic disorder through the larger reorganization energy, reducing the spatial extent of the exciton and suppressing overlap between states. This behavior is reflected in the decay of the spatial overlap and the corresponding increase in the inverse participation factor $1/N_{\mathrm{eff}}$. In the FE-dominated regime ($\bar{f}_{\mathrm{CT}} \to 0$), transitions are limited by weak coupling to environmental fluctuations, whereas strong localization suppresses transport in the CT-dominated limit ($\bar{f}_{\mathrm{CT}} \to 1$). These opposing rate-limiting mechanisms lead to a turnover in the transfer rate, with a maximum at intermediate $\bar{f}_{\mathrm{CT}}$. The results in Fig.~\ref{fig:turnover} predict that an intermediate degree of FE--CT mixing ($\bar{f}_{\mathrm{CT}} \sim 0.25$) optimizes exciton transport.

The turnover behavior of the microscopic exciton transfer rate is connected to the experimentally observed quenching dynamics through a minimal kinetic model describing ultrafast hole transfer into an interfacial manifold followed by a slower recovery process (see Fig.~\ref{fig:tascheme}). The population dynamics can be described by the bulk exciton population $N_X(t)$
\begin{align}
\frac{dN_X}{dt} &= -R_{\mathrm{mig}}\, N_X, \label{eq:NX}
\end{align}
and the quenched population $N_Q(t)$
\begin{align}
\frac{dN_Q}{dt} &= R_{\mathrm{mig}}\, N_X - R_{\mathrm{CR}}\, N_Q, \label{eq:NQ}
\end{align}
where $R_{\mathrm{mig}}$ is the exciton migration rate and $R_{\mathrm{CR}}$ is the charge-recombination (recovery) rate. The migration rate is assumed to be governed by the same microscopic exciton transfer processes captured by the Redfield model, such that $R_{\mathrm{mig}} \propto R_{\mathrm{ET}}(\bar{f}_{\mathrm{CT}})$. For the initial conditions $N_X(0)=N_0$ and $N_Q(0)=0$, the bulk exciton population decays as
\begin{equation}
N_X(t)=N_0 \exp\left(-R_{\mathrm{mig}} t\right),
\label{eq:NXsol}
\end{equation}
and the quenched population evolves according to
\begin{equation}
\begin{split}
N_Q(t)
&=
N_0
\frac{R_{\mathrm{mig}}}
{R_{\mathrm{CR}}-R_{\mathrm{mig}}}
\\
&\qquad\times
\left[
\exp\left(-R_{\mathrm{mig}} t\right)
-
\exp\left(-R_{\mathrm{CR}} t\right)
\right],
\end{split}
\label{eq:NQsol}
\end{equation}
for $R_{\mathrm{CR}} \neq R_{\mathrm{mig}}$. This expression exhibits a characteristic rise and decay, reflecting the competition between population transfer at rate $R_{\mathrm{mig}}$ and loss through charge recombination at rate $R_{\mathrm{CR}}$. 

In this description, the hole transfer dynamics are controlled by $R_{\mathrm{mig}}$, while $R_{\mathrm{CR}}$ captures the slower recovery process that is not central to the transport mechanism. Because $R_{\mathrm{mig}}$ is proportional to the microscopic exciton transfer rate $R_{\mathrm{ET}}(\bar{f}_{\mathrm{CT}})$ obtained from Redfield theory, the experimentally observed quenching dynamics provide a constraint on the degree of FE--CT mixing. Within this model, the peak of the transient absorption response occurs at
\begin{equation}
t_{\mathrm{peak}}=
\frac{\ln\!\left(R_{\mathrm{CR}}/R_{\mathrm{mig}}\right)}
{R_{\mathrm{CR}}-R_{\mathrm{mig}}}.
\label{eq:tpeak}
\end{equation}
Using the peak times extracted from films with 20 and 40 nm thicknesses, which exhibit the clearest quenching signatures, we obtain $t_{\mathrm{peak}} \approx  12.0 \pm 0.4$ ps. Taking the recovery time from the same data sets gives a charge-recombination time of $580.6 \pm 14.1~\mathrm{ps}$, or $R_{\mathrm{CR}} = 1.72 \times 10^{-3}~\mathrm{ps}^{-1}$. Substituting these values into Eq.~\eqref{eq:tpeak} results in an effective migration rate $R_{\mathrm{mig}} \approx 0.47~\mathrm{ps}^{-1}$, corresponding to a characteristic transport time of $\sim 2.1~\mathrm{ps}$. This value is consistent with rapid transport within the few-nanometer interfacial capture region inferred from the thickness dependence of the quenching amplitude (see Eq.~\eqref{eq:quench_depth}).

The empirical migration rate $R_{\mathrm{mig}}$ can be related to the microscopic Redfield rate $R_{\mathrm{ET}}$ by considering a kinetic model in which transport to the interfacial capture region proceeds through $N_{\mathrm{step}}$ sequential exciton-transfer events. The mean migration time is then given by the sum of the individual transfer times, $\langle t \rangle \approx N_{\mathrm{step}}/R_{\mathrm{ET}}$, leading to
\begin{equation}
R_{\mathrm{mig}} \equiv \frac{1}{\langle t \rangle} = \frac{R_{\mathrm{ET}}}{N_{\mathrm{step}}}.
\end{equation}
In contrast, for diffusive hopping with frequent back-transfer, the mean first-passage time grows quadratically with distance, giving $R_{\mathrm{mig}} \propto R_{\mathrm{ET}}/N_{\mathrm{step}}^2$. The experimentally observed ratio between the microscopic transfer rate ($R_{\mathrm{ET}} \sim 1$--3 ps$^{-1}$ in Fig.~\ref{fig:turnover}(a)) and the effective migration rate ($R_{\mathrm{mig}} \sim 0.5~\mathrm{ps}^{-1}$) corresponds to $N_{\mathrm{step}} \sim 2$--6, indicating that transport to the interface occurs through a small number of exciton-transfer events. Furthermore, the measured quenching times correspond to mean values of $\bar{f}_{\mathrm{CT}}$ less than 50\%, as shown in Fig.~\ref{fig:turnover}.

Independent support for the existence of a manifold of closely spaced hybrid FE--CT exciton states is provided by the transient absorption anisotropy measurements of Y6 in solution and films presented in the Supplemental Material \cite{SM_Y6}. Although the linear absorption spectra of the films differ substantially from those of the solution \cite{zouInsightExcitationStates2020,kashaniRelatingReorganizationEnergies2023}, the individual exciton resonances remain unresolved, motivating the present nonlinear optical approach in which transition dipole orientations provide sensitivity to the underlying electronic structure. With the solution serving as a molecular reference, the pronounced reduction of the film anisotropy indicates that multiple optically active exciton states with distinct transition dipole orientations contribute within the excitation bandwidth at 810~nm. These findings are also consistent with transient absorption anisotropy measurements previously reported for Y6 films \cite{saynerExcitonDiffusionLow2024}, indicating that the presence of multiple hybrid FE--CT exciton states is a robust feature of Y6. This observation supports the the hybrid FE--CT manifold assumed in the present model and supports the interpretation that interfacial quenching proceeds through successive exciton transfer events within the hybrid FE--CT manifold prior to hole transfer.

The present analysis is consistent with recent theoretical and spectroscopic studies of Y6, which report hybrid FE--CT excitations with significant but not dominant CT character \cite{fujitaCoherentDynamicsMixed2016,gianniniRoleChargeTransfer2024,cerdáTuningExcitonDiffusion2025}. In these systems, CT states lie energetically close to Frenkel excitons, forming a dense manifold of hybrid FE--CT exciton states that governs the excited-state dynamics. As the CT character increases, localization of the eigenstates reduces their spatial overlap, thereby suppressing the exciton transfer rate. The turnover predicted by the present Redfield model implies that the relevant excitations possess an intermediate CT character, typically below $\sim 50\%$, where exciton transport is enhanced without incurring strong localization. 

\section{Conclusions}

In summary, we have investigated exciton transport and quenching dynamics in Y6 films using ultrafast transient absorption spectroscopy in combination with a reduced theoretical model. The experimentally extracted migration rate corresponds to a characteristic transport time of $\sim 2$ ps, indicating rapid motion of excitations near the interfacial capture region. Prior to hole transfer at the interface, our data suggest that exciton migration occurs over a length scale of $\sim4$ nm and is inconsistent with long-range diffusion. Instead, the observed dynamics indicate a delocalization-mediated transport mechanism involving a small number of sequential exciton transfer events. Fitting the absorption lineshape constrains the parameters of the model Hamiltonian that governs the microscopic exciton transfer rates, yielding an intermediate CT admixture and reorganization energies consistent with strong FE--CT coupling. The analysis further indicates that spectral broadening at higher energies primarily arises from CT intensity borrowing rather than superradiant H- or J-type aggregation. Overall, the analysis suggests that the underlying exciton states possess intermediate CT character ($\bar{f}_{\mathrm{CT}} \lesssim 50\%$), for which the microscopic exciton transfer rate is maximized near the turnover shown in Fig.~\ref{fig:turnover}.

More broadly, the approach developed here provides a general framework for characterizing exciton dynamics in non-fullerene acceptors and related molecular semiconductors. The reduced Hamiltonian facilitates the systematic extraction of key physical parameters governing light absorption and exciton transport from experimental spectroscopic data. While the present work focuses on steady-state absorption and pump--probe kinetics, the model can be generalized to describe other spectroscopic techniques, including time-resolved photoluminescence and two-dimensional electronic spectroscopy. This versatility provides a common physical basis for relating optical responses to exciton FE--CT character and transport mechanisms across diverse molecular semiconductors.

\begin{acknowledgments}
This work is supported by the National Science Foundation under Grant No.~CHE-2247159 (S.M., Z.G., and A.M.) and the Office of Naval Research under Award N00014-24-1-2107 (J.O., Y.L., and W.Y.). This work was performed in part at the Chapel Hill Analytical and Nanofabrication Laboratory (CHANL), a member of the North Carolina Research Triangle Nanotechnology Network (RTNN), which is supported by the National Science Foundation under Grant No.~ECCS-2025064, as part of the National Nanotechnology Coordinated Infrastructure (NNCI).
\end{acknowledgments}

\section*{Data Availability}

The data that support the findings of this study are available from the corresponding author upon reasonable request.

\appendix
\section{Decomposition of the Overlap Factor}
\label{app:overlap}

In this appendix we justify the approximation used to obtain the factorized rate expression in Eq.~\eqref{eq:k_factorized_main} from the general Redfield expression in Eq.~\eqref{eq:redfield_decomposition_main} by analyzing the structure of the overlap factor and defining an effective delocalization size. The exact overlap factor entering the Redfield rate is
\begin{equation}
\mathcal O_{\alpha\beta}
=
\sum_n
|C_n^{(\alpha)}|^2
|C_n^{(\beta)}|^2
+
\sum_{n<m}
|D_{nm}^{(\alpha)}|^2
|D_{nm}^{(\beta)}|^2,
\end{equation}
which depends on both the spatial extent of each eigenstate and their relative location.
To make this dependence more transparent, we approximate the overlap as
\begin{equation}
\mathcal O_{\alpha\beta}
\approx
P_{\alpha\beta}
\frac{1}{\sqrt{N_{\mathrm{eff}}^{(\alpha)}N_{\mathrm{eff}}^{(\beta)}}}.
\end{equation}
Here, $N_{\mathrm{eff}}^{(\alpha)}$ characterizes the number of diabatic configurations contributing to eigenstate $|\alpha\rangle$,
\begin{equation}
N_{\mathrm{eff}}^{(\alpha)}
=
\frac{1}{
\displaystyle
\sum_n |C_n^{(\alpha)}|^4
+
\sum_{n<m} |D_{nm}^{(\alpha)}|^4
},
\end{equation}
which is the inverse participation ratio commonly used to quantify localization. Large $N_{\mathrm{eff}}^{(\alpha)}$ corresponds to delocalized eigenstates, whereas small $N_{\mathrm{eff}}^{(\alpha)}$ indicates stronger localization.

To quantify the geometric contribution to the overlap, we project the eigenstates onto molecular sites. The weight of eigenstate $\alpha$ on site $n$ is
\begin{equation}
w_{n\alpha}
=
|C_n^{(\alpha)}|^2
+
\sum_{m>n} |D_{nm}^{(\alpha)}|^2
+
\sum_{m<n} |D_{mn}^{(\alpha)}|^2 .
\end{equation}
The normalized distribution is $\tilde w_{n\alpha}=w_{n\alpha}/\sum_m w_{m\alpha}$, and the spatial overlap is defined as
\begin{equation}
P_{\alpha\beta}
=
\sum_n \sqrt{\tilde w_{n\alpha}\tilde w_{n\beta}}.
\end{equation}
This quantity approaches unity when the two eigenstates occupy the same spatial region and vanishes when they are localized on disjoint regions. 

Combining this approximation with the separation of spectral density contributions in Eq.~\eqref{eq:redfield_decomposition_main}, we approximate the weighted sum over diabatic configurations as
\begin{equation}
\begin{aligned}
&
S_{\mathrm{FE}}(\omega_{\alpha\beta})
\sum_n
|C_n^{(\alpha)}|^2
|C_n^{(\beta)}|^2
\\
&\qquad+
S_{\mathrm{CT}}(\omega_{\alpha\beta})
\sum_{n<m}
|D_{nm}^{(\alpha)}|^2
|D_{nm}^{(\beta)}|^2
\\
&\approx
S_{\mathrm{eff}}(\omega_{\alpha\beta};\bar{f}_{\mathrm{CT}})
\Biggl[
\sum_n
|C_n^{(\alpha)}|^2
|C_n^{(\beta)}|^2
\\
&\qquad\qquad+
\sum_{n<m}
|D_{nm}^{(\alpha)}|^2
|D_{nm}^{(\beta)}|^2
\Biggr].
\end{aligned}
\end{equation}
Using the definition of the overlap factor,
$\mathcal O_{\alpha\beta}$, Eq.~\eqref{eq:redfield_decomposition_main} reduces to the factorized exciton transfer rate given in Eq.~\eqref{eq:k_factorized_main},
\begin{equation}
R_{\alpha\beta}
\approx
\frac{1}{\hbar^2}
S_{\mathrm{eff}}(\omega_{\alpha\beta};\bar{f}_{\mathrm{CT}})
\,\mathcal O_{\alpha\beta}.
\end{equation}

\section{Disorder-Dependent Localization and Spatial Overlap}
\label{app:localization}

In this appendix we relate the strength of energetic disorder to the spatial extent of exciton eigenstates. In disordered tight-binding models, a commonly used weak-disorder estimate for the localization length is
\begin{equation}
\xi\!\left(\sigma_{\mathrm{eff}}\right)
\sim
\left(
\frac{J_{\mathrm{eff}}}
{\sigma_{\mathrm{eff}}}
\right)^2,
\end{equation}
which is consistent with perturbative treatments of Anderson localization in one dimension \cite{andersonAbsenceDiffusionCertain1958,kramerLocalizationTheoryExperiment1993}. Here, the effective disorder $\sigma_{\mathrm{eff}}$ is defined in Eq.~\eqref{eq:sigmaeff_main}, and the effective electronic coupling is approximated by
\begin{equation}
J_{\mathrm{eff}}(\bar{f}_{\mathrm{CT}})
=
(1-\bar{f}_{\mathrm{CT}})J_{\mathrm{FE}}
+
\bar{f}_{\mathrm{CT}}J_{\mathrm{CT}},
\end{equation}
which linearly interpolates between the FE and CT coupling strengths according to the average CT fraction. Increasing the reorganization energy enhances the effective energetic disorder and consequently reduces the localization length. The spatial overlap between two eigenstates separated by a characteristic distance $r_0$ can be approximated by an exponential tail overlap,
\begin{equation}
P(\bar{f}_{\mathrm{CT}})
=
\exp\!\left[
-\frac{r_0}
{\xi\!\left(\sigma_{\mathrm{eff}}(\bar{f}_{\mathrm{CT}})\right)}
\right],
\label{eq:P_fct}
\end{equation}
which decreases as increasing CT character enhances the effective disorder and reduces the localization length. 

To connect this behavior with the present reduced description, we express the overlap as the product of the spatial overlap factor and the inverse participation factor,
\begin{equation}
\mathcal O(\bar{f}_{\mathrm{CT}})
\approx
\frac{
P(\bar{f}_{\mathrm{CT}})
}
{
N_{\mathrm{eff}}\!\left(\sigma_{\mathrm{eff}}(\bar{f}_{\mathrm{CT}})\right)
}.
\label{eq:overlap_fct_compact}
\end{equation}
Equations~\eqref{eq:overlap_fct_compact} and \eqref{eq:Neff_main} can then be combined to obtain
\begin{equation}
\begin{split}
\mathcal O(\bar{f}_{\mathrm{CT}})
&\approx
\exp\!\left[
-\frac{r_0}
{\xi\!\left(\sigma_{\mathrm{eff}}(\bar{f}_{\mathrm{CT}})\right)}
\right]
\\
&\qquad\times
\left[
N_{\min}
+
\frac{N_{\max}-N_{\min}}
{1+\left(\sigma_{\mathrm{eff}}(\bar{f}_{\mathrm{CT}})/\sigma_c\right)^2}
\right]^{-1}.
\end{split}
\label{eq:overlap_fct_explicit}
\end{equation}
This expression explicitly shows that increasing CT character suppresses the overlap through the combined effects of disorder-induced localization and the inverse participation factor, thereby providing the microscopic origin of the turnover in the Redfield transfer rate.

\bibliographystyle{apsrev4-2}
\bibliography{ZoteroLibraryAbbrev}
\end{document}